\documentclass[aps,prl,twocolumn,superscriptaddress,showpacs,amsmath,amssymb,floatfix]{revtex4-1}
\pdfoutput=1
\usepackage{color}
\usepackage{amssymb}
\usepackage{amsmath}
\usepackage{mathtools}
\usepackage{bm}
\usepackage{url}
\usepackage{hyperref}
\usepackage{graphicx}
\usepackage{soul}
\usepackage{dcolumn}
\usepackage{xcolor}

\begin{document}

\title{Optimal navigation strategies for microswimmers on curved manifolds}
\author{Lorenzo Piro}
\affiliation{Max Planck Institute for Dynamics and Self-Organization (MPIDS), 37077 G{\"o}ttingen, Germany}
\author{Evelyn Tang}
\affiliation{Max Planck Institute for Dynamics and Self-Organization (MPIDS), 37077 G{\"o}ttingen, Germany}
\author{Ramin Golestanian}
\email{ramin.golestanian@ds.mpg.de}
\affiliation{Max Planck Institute for Dynamics and Self-Organization (MPIDS), 37077 G{\"o}ttingen, Germany}
\affiliation{Rudolf Peierls Centre of Theoretical Physics, University of Oxford, Oxford OX1 3PU, United Kingdom}

\date{\today}

\begin{abstract}
Finding the fastest path to a desired destination is a vitally important task for microorganisms moving in a fluid flow. We study this problem by building an analytical formalism for overdamped microswimmers on curved manifolds and arbitrary flows. We show that the solution corresponds to the geodesics of a Randers metric, which is an asymmetric Finsler metric that reflects the irreversible character of the problem. Using the example of a spherical surface, we demonstrate that the swimmer performance that follows this ``Randers policy’’ always beats a more direct policy. A study of the shape of isochrones reveals features such as self-intersections, cusps, and abrupt nonlinear effects. Our work provides a link between microswimmer physics and geodesics in generalizations of general relativity. 
\end{abstract}

\maketitle

\paragraph{Introduction.---}

It is beneficial for microorganisms such as bacteria, algae, or spermatozoa to employ sensing mechanisms equipped with adaptation strategies to control their motility machinery in order to find the fastest path towards a desired destination \cite{berg}, e.g. when tracking a food source \cite{Bray2000} or seeking light \cite{Bennett2015}. Such navigation typically takes place in the presence of a fluid flow or an external force landscape, which can hinder or help their motion. The optimal path is hence distinct from the shortest path, rendering this a complex problem in the field of active matter \cite{Gompper2020}. In addition, artificial micro- and nanoswimmers \cite{Golestanian2007} with active external controls (e.g. via chemical \cite{RG-phoretic,stark} and electromagnetic fields \cite{mano,Tierno2008b}, feedback loops \cite{bauerle, khadka}, and geometric features of boundaries \cite{Das2015}) can increasingly be engineered to execute specialized tasks in complex environments. These have crucial technological and medical applications ranging from targeted delivery of drugs \cite{park}, genes \cite{qiu}, or other cargo \cite{demirors}, to prevention of dental biofilm \cite{Villa2020}. 



Optimal navigation was first addressed by Zermelo, who studied a ship navigating in the presence of an external wind \cite{zermelo}. Recent work has explored this problem for microorganisms navigating on 2D surfaces for a limited class of force fields featuring specific symmetries (constant, 1D-, shear- or vortex-fields) \cite{liebchen}. Other approaches include algorithmic optimization procedures based on the application of machine learning to active motion \cite{colabrese,biferale,schneider}. Here, we aim to develop an analytical formalism for optimal navigation in an over-damped system, which can be used on curved manifolds and arbitrary stationary flows. Adopting recent mathematical results from differential geometry \cite{shen,bao}, we show that this problem can be mapped onto geodesics of a Finsler-type geometry \cite{Finsler} with a Randers metric \cite{randers}. Finsler spaces have been used to construct geometric descriptions in many areas of physics, with applications ranging from electron motion in magnetic flows \cite{gibbons} to quantum control \cite{brody} and test theories of relativity \cite{Golestanian1995}. The particular choice of the asymmetric Randers metric allows us to characterize the irreversibility of the optimal trajectory in this non-equilibrium problem. 

We start by illustrating the formalism and discussing some general properties of the system. Then, we apply these concepts to a specific setup and study how following Randers geodesics can reduce the travel time to reach a target compared to when the microswimmer heads constantly towards it. Lastly, we analyze the \emph{isochrones}---curves of equal travel time---to investigate more generally the shape of optimal paths.

\begin{figure*}[!t]
\centering
\includegraphics[width=0.9\textwidth]{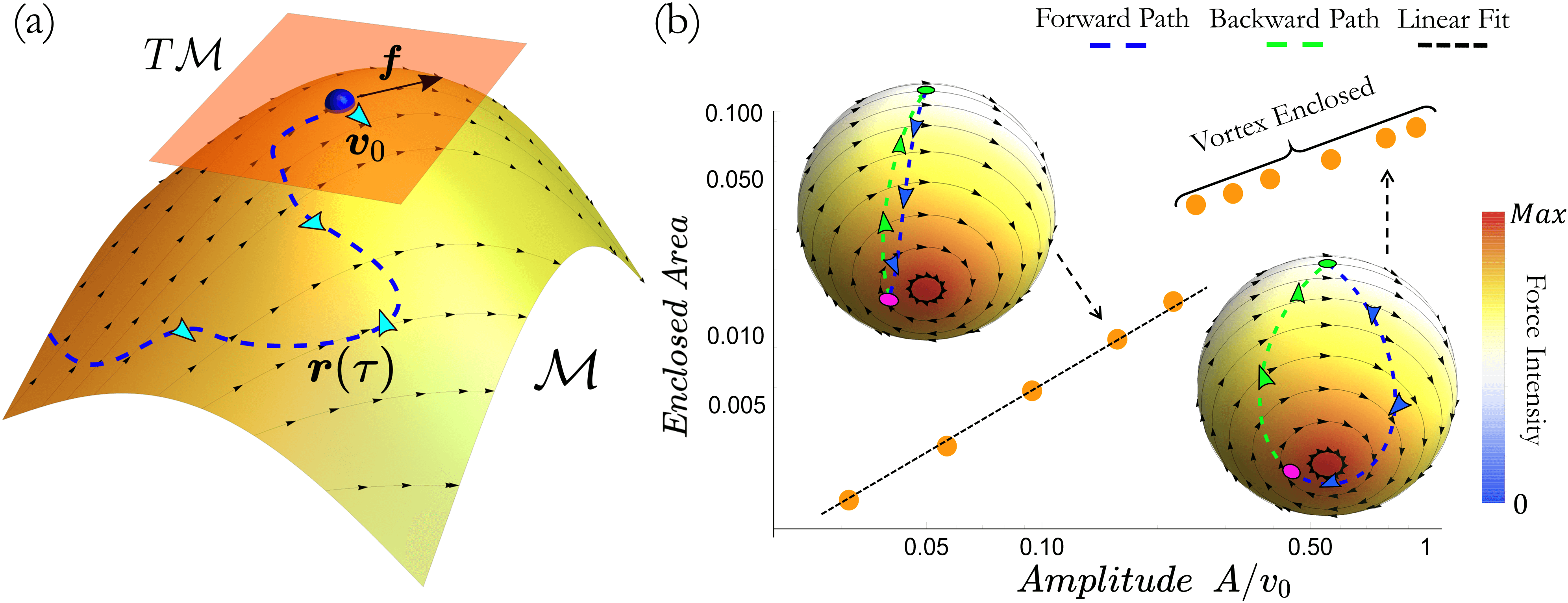}
\caption{(a) Microswimmer trajectory ${\bm r}(\tau)$ (blue dashed line) on a Riemannian manifold $\mathcal{M}$. The microswimmer moves in the tangent space $T\mathcal{M}$ under the influence of the force $\bm f$ (black arrows) and its self-propelling velocity ${\bm v}_0$ which is marked with cyan arrows. (b) Fraction of a spherical surface enclosed in the optimal forward-backward loop as a function of the force amplitude. There is a jump when the vortex is enclosed in the loop, also seen in the change in the scaling behavior, which is linear (black dashing) for small values of the force amplitude. There are two exemplar trajectories for given values of the force amplitude as indicated by dashed black arrows. The forward (blue dashing) and backward (green dashing) paths connect the following two points: $(\theta_0,\phi_0)=(\frac{\pi}{2},0) \rightleftarrows (\theta_1,\phi_1)=(\frac{11\pi}{12},\frac{14\pi}{10})$. The arrows on top of each trajectory indicate the heading direction of the microswimmer. The intensity (color) gradient on each sphere shows the force intensity, while the solid black arrows show the force direction.
  \label{fig1area}}
\vspace{-0.5truecm}
\end{figure*}

\paragraph{Curved manifolds and Finsler geometry.---}Consider a microswimmer that is free to move on a smooth Riemannian manifold $\mathcal{M}$ [see Fig. \ref{fig1area}(a)] equipped with a positive definite metric $h$, such that the corresponding norm of any tangent vector ${\bm x}\in T\mathcal{M}$ can be calculated via $|{\bm x}|_h^2=h_{ij} x^i x^j$, where Einstein's summation convention is used \cite{schutz}. The Riemannian metric can also be used to define the scalar product of any two tangent vectors ${\bm x},{\bm y}\in T\mathcal{M}$ as $h_{ij}x^iy^j$. Let us now assume that the microswimmer moves on such a surface with the self-propulsion velocity ${\bm v}_0$, which corresponds to a constant speed $|{\bm v}_0|_h \equiv v_0$. 
The motion takes place in the presence of a time-independent force field $\bm f({\bm r})$, which may in general include a contribution due to advection by the solvent flow velocity (note that the friction coefficient is set to unity). 
The over-damped motion of the microswimmer can therefore be described as follows
\begin{equation}
	 \frac{d{\bm r}}{d\tau}={\bm v}_0+{\bm f}({\bm r}(\tau)) \ ,
	\label{eq:eqmotion}
\end{equation}
where  $\tau$ is the swimmer (``proper’’) time [see Fig. \ref{fig1area}(a)]. We neglect rotational noise and assume full control over the direction of microswimmer propulsion as described by ${\bm v}_0$. This means that the direction of propulsion is steered by a protocol that selects the appropriate active angular velocity to make it follow a prescribed path.


To show how Finsler geometry enters the optimal navigation problem on curved manifolds, we consider the time for a microswimmer to go from one point ${\bm r}_{\rm A}$ to another ${\bm r}_{\rm B}$ on the surface via the trajectory ${\bm r}(s)$ that is parametrized with $s$:
\begin{equation}
	T=\int_{{\bm r}_{\rm A}}^{{\bm r}_{\rm B}} d\tau = \int_{{\bm r}_{\rm A}}^{{\bm r}_{\rm B}} \frac{d s}{v}\equiv  \int_{{\bm r}_{\rm A}}^{{\bm r}_{\rm B}} d s\, \mathcal{L}[s,{\bm r}(s),\dot{\bm r}(s)]\ ,
	\label{eq:time}
\end{equation}
where $v\equiv \frac{ds}{d\tau}$, $\dot{\bm r}\equiv \tfrac{d{\bm r}}{ds}$, and the Lagrangian $\mathcal{L} \equiv v^{-1}$ is defined by identifying the traveling time as an action. Using Eq. \eqref{eq:eqmotion}, we obtain the following expression for the Lagrangian
\begin{equation}
	\mathcal{L}=\sqrt{a_{ij}\dot{r}^i\dot{r}^j}+b_i\dot{r}^i \ .
	\label{eq:lagr2}
\end{equation}
where we have used the definitions $a_{ij}\equiv h_{ij}\lambda+f_i f_j \lambda^2$, $b_i \equiv -f_i \lambda$, and $\lambda^{-1} \equiv v_0^2-h_{ij} f^i f^j$, with $f_i=h_{ij} f^j$. We now make the observation that the resulting Lagrangian has all the defining features to be a Finsler metric of Randers type \cite{randers}, if and only if the condition $|{\bm f}|_h< v_0$ is fulfilled
at any point on the surface and any time; see Ref. \cite{bao} for a proof. This implies that the self-propulsion is assumed to be able to overpower the external force ${\bm f}$ at all points. Such a constraint ensures that $\mathcal{L}$ is strongly convex and positive-definite, which are two necessary features for identification as a Randers metric \cite{bao2,chern,cheng}. 

\paragraph*{Randers spaces and irreversibility.---}

\begin{figure*}[!t]
\centering
\includegraphics[width=0.9\textwidth]{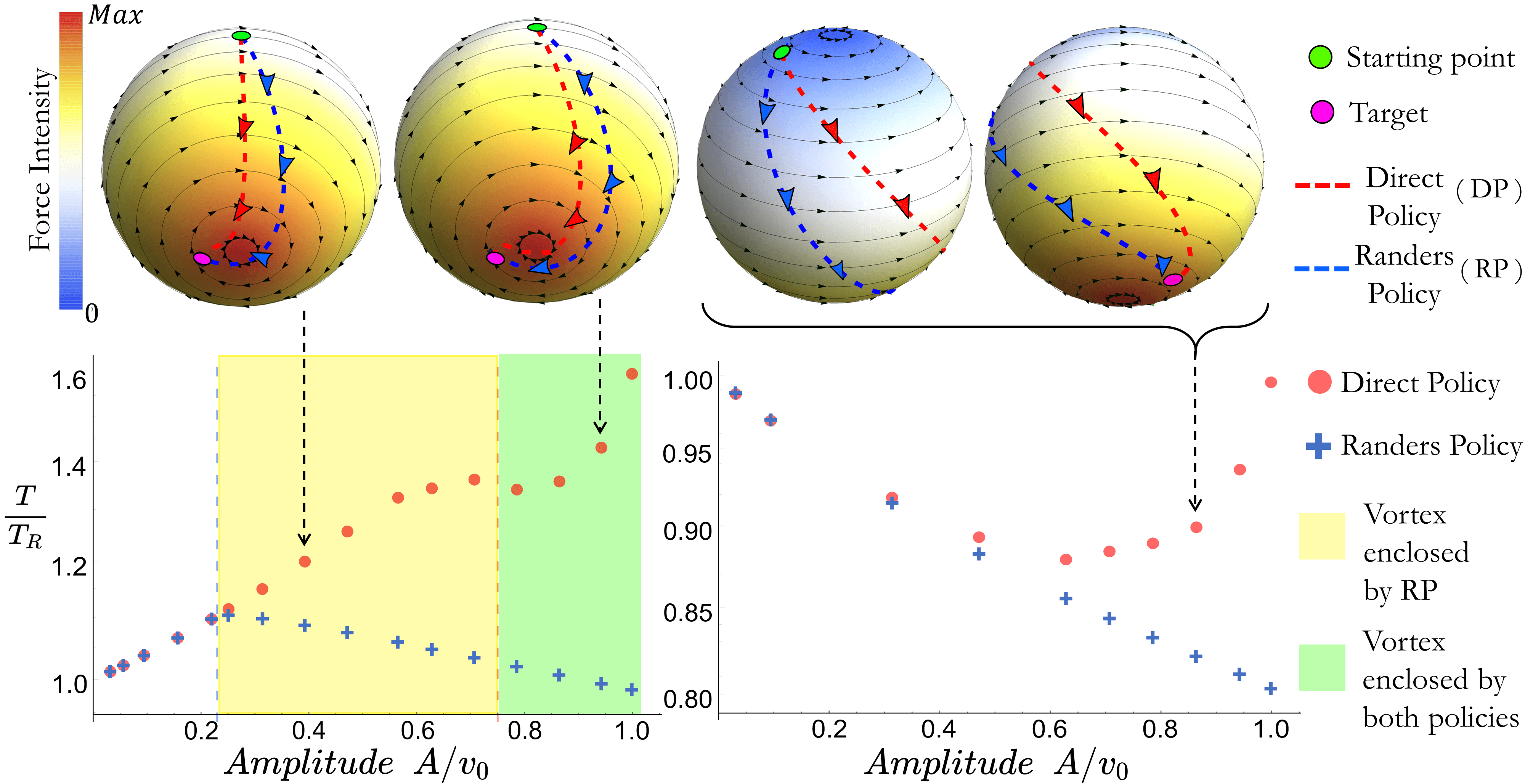} 
\caption{Comparison between the Randers Policy (RP, blue crosses) and the Direct Policy (DP, red circles) in terms of the arrival times $T$ in units of $T_R$---the optimal navigation time in the absence of $\bm f$---as a function of the force amplitude. For both strategies, the target counts as reached when the microswimmer enters a region of radius $\epsilon=0.01$. On the left: analysis of the paths connecting the points $(\theta_0,\phi_0)=(\frac{\pi}{2},0)\to(\theta_1,\phi_1)=(\frac{11\pi}{12},\frac{14\pi}{10})$. On the right: study of the paths linking two diametrically opposite points $(\theta_0,\phi_0)=(\frac{\pi}{8},0)\to(\theta_1,\phi_1)=(\frac{7\pi}{8},\pi)$. 
In either case the DP is sub-optimal and there is a clear gain in following the RP, especially when the force is stronger. In the upper part of the figure there are three different exemplar trajectories for given values of the force amplitude as indicated by the dashed black arrow. The dashed lines on every sphere represent the paths for each navigation strategy (RP: blue; DP: red) and the arrows on top of them show the corresponding heading direction of the microswimmer. The gradient on each sphere indicates the force intensity, while the solid black arrows represent its direction.
\label{figDP}}
\vspace{-0.5truecm}
\end{figure*}

Randers spaces are often referred to as a special class of non-reversible Finsler spaces \cite{bao2}. This is due to the presence of the second term in \eqref{eq:lagr2}, namely $b_i\dot{r}^i$, which makes the metric tensor manifestly asymmetric under time reversal, i.e. $\mathcal{L}(\dot{r}^i)\neq\mathcal{L}(-\dot{r}^i)$. Due to this asymmetry, in presence of an external force the optimal forward path (between ${\bm r}_{\rm A}$ and ${\bm r}_{\rm B}$) will in general be different from the backward one (${\bm r}_{\rm B}$ to ${\bm r}_{\rm A}$). In other words, the optimal backward path is distinct from the the time-reversed forward path, which highlights the out-of-equilibrium character of the navigation problem we study. In contrast, Riemannian geodesics (in the absence of any external force) are reversible since the corresponding metric tensor $h$ is symmetric \cite{lee}. This property of Randers metrics is illustrated with a concrete example in Fig. \ref{fig1area}(b) and studied in more detail below.


Since $\mathcal{L}$ is a homogeneous function of degree one with respect to $\dot{r}^i$, we can introduce the {\em fundamental tensor} 
\begin{equation}
	g_{ij}\equiv\frac12 \frac{\partial^2 \mathcal{L}^2}{\partial{\dot{r}^i}\partial{\dot{r}^j}} \ ,
	\label{eq:g_ij}
\end{equation}
which is also positive definite due to the convexity condition \cite{cheng}. 
For the Randers metric of Eq. \eqref{eq:lagr2}, we find
\begin{equation}
g_{ij}=\left(1+\frac{b_i \dot{r}^i}{\sqrt{a_{ij}\dot{r}^i\dot{r}^j}}\right) \left(a_{ij}-\ell_i \ell_j\right)+\left(b_i+\ell_i\right) \left(b_j+\ell_j\right)\ ,
	\label{eq:g_ij-2}
\end{equation}
where $\ell_i \equiv a_{ij}\dot{r}^j/\sqrt{a_{ij}\dot{r}^i\dot{r}^j}$. 
In order to determine the time-minimizing paths, we solve the Euler-Lagrange equations for the corresponding energy functional ${\cal E}=\frac{1}{2}\mathcal{L}^2=g_{ij}\dot{r}^i\dot{r}^j$, namely $\frac{d}{ds}\big(\frac{\partial {\cal E}}{\partial \dot{r}^m}\big)=\frac{\partial {\cal E}}{\partial r^m}$. The paths minimizing $\int ds \,{\cal E}$, which also minimize the travel time $T$, satisfy the Randers metric geodesic equation
\begin{equation}
	\ddot{r}^k+\Gamma^k_{ij}\dot{r}^i\dot{r}^j=0 \ ,
	\label{eq:geodesic}
\end{equation}
where the Christoffel symbol $\Gamma^k_{ij}$ is defined via $\Gamma^k_{ij}\equiv\frac{1}{2}g^{km}(g_{im,j}+g_{jm,i}-g_{ij,m})$, with $g^{km}$ being the inverse of the fundamental tensor defined in \eqref{eq:g_ij} and $g_{ij,m}\equiv\partial_{m}g_{ij}$. Thus, the solutions of the geodesic equation \eqref{eq:geodesic} provide optimal navigation paths for a microswimmer moving in the presence of the force field ${\bm f}$ on a generic Riemannian manifold $\mathcal{M}$. In what follows, we apply these theoretical concepts to the case in which the motion takes place on a sphere.


\paragraph{Optimal navigation on a sphere.---}

Let us consider a sphere of radius unity embedded in $\mathbb{R}^3$. The position of the microswimmer on this surface can be written in spherical coordinates as ${\bm r}=(\theta,\phi)$. 
The corresponding Riemannian metric $h$ in spherical coordinates has the components $h_{\theta \theta}=1$, $h_{\phi \phi}=\sin^2\theta$, and $h_{\theta \phi}=h_{\phi \theta}=0$.
The force field is then ${\bm f}({\bm r})=f_\theta(\theta,\phi)\hat{\bm e}_\theta+f_\phi(\theta,\phi)\hat{\bm e}_\phi$. As an example, we choose $f_\theta(\theta,\phi)=0$ and $f_\phi(\theta,\phi)=\frac{A \theta}{\pi \sin\theta}$, 
where $A$ sets the amplitude of the field, which is constrained as  $A<v_0$. This divergence-free force field is characterized by a pair of vortices at the poles of the sphere and its intensity is maximum (minimum) at the south (north) pole. 

We can then write the explicit expression of the Randers metrics $\mathcal{L}$ in our case as follows
\begin{equation*}
    \mathcal{L}=\frac{\sqrt{v_0^2 \sin^2{\theta} \dot{\phi}^2+(v_0^2-A^2 \theta^2/\pi^2)\dot{\theta}^2}- A \dot{\phi} \theta \sin{\theta}/\pi}{v_0^2-A^2 \theta^2/\pi^2} \ .
\end{equation*}
It is then possible to determine the fundamental tensor $g_{ij}$, the relative Christoffel symbols $\Gamma^{k}_{ij}$ and the corresponding geodesic equations using their definitions in \eqref{eq:g_ij} and \eqref{eq:geodesic}. 
We further choose the following initial conditions: $\theta(0)=\theta_0$, $\phi(0)=\phi_0$, $\dot{\theta}(0)=-\sin\varphi_0+f_{\theta}(\theta_0,\phi_0)$, and $\dot{\phi}(0)=\frac{\cos\varphi_0}{\sin\theta_0}+f_{\phi}(\theta_0,\phi_0)$.
Here, $(\theta_0,\phi_0)$ is the starting position while $\varphi_0$ represents the initial heading direction of the microswimmer (measured counterclockwise with respect to the $\hat{\bm e}_\phi$ direction), which we scan when using the shooting method, selecting the one that takes the shortest time. 
Moreover, we parametrize the trajectory using the proper time of the microswimmer (i.e. we set $s=v_0 \tau$), which implies that $\mathcal{L}$ will be a conserved quantity along these paths. 

\begin{figure*}[!t]
\centering
\includegraphics[width=0.9\textwidth]{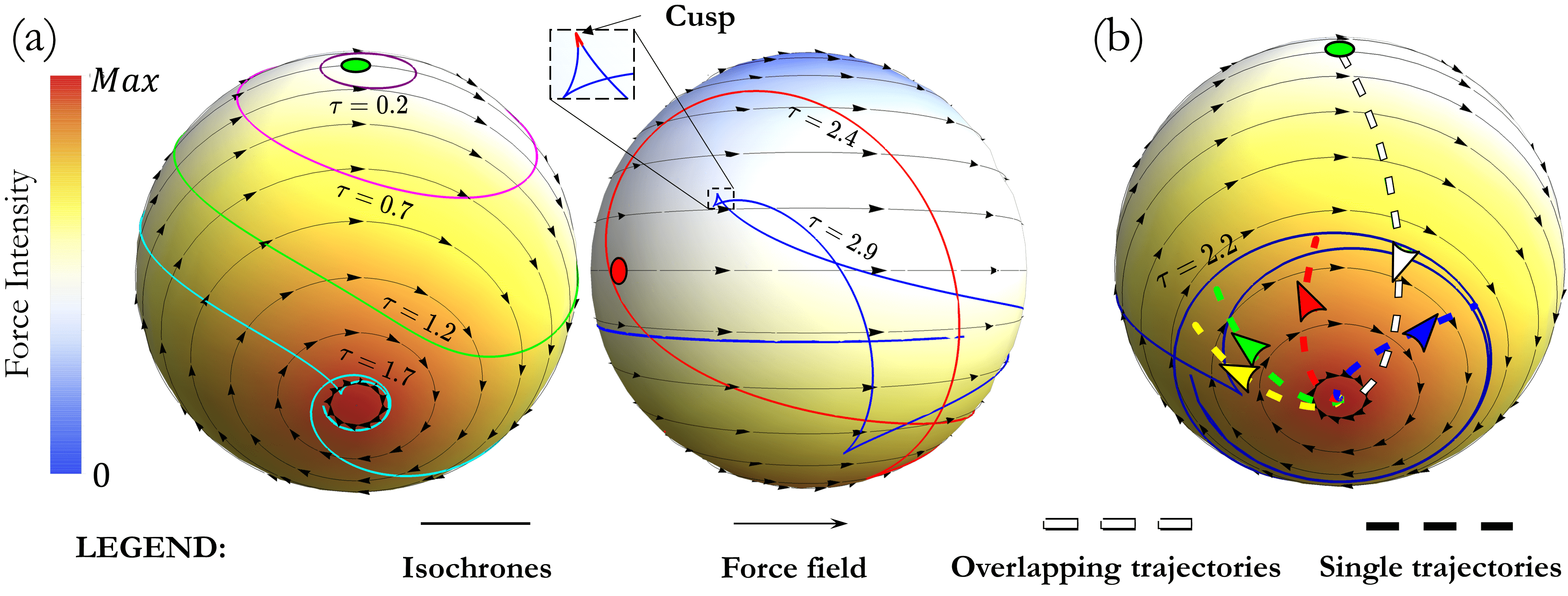} 
\caption{Analysis of the isochrones starting from the point $(\theta_0,\phi_0)=(\pi/2,0)$ (green circle) in the presence of the force field with $A=\frac{1}{2}v_0$. The red circle indicates the point diametrically opposite to the starting point and provides a guide to the eye. The color code on each sphere shows the force intensity from small (blue) to high (red), while the black arrows represent its direction. The time $\tau$ is reported in units of $v_0^{-1}$. (a) Isochrones (solid lines) at six different times. On the left: view of a region of stronger force. On the right: front view with an enlargement of a cusp highlighted in red, a point with a high density of geodesics. Also, notice the presence of self-intersections, points where two optimal paths collide. (b) Representation of 4 specific paths (dashed lines) passing close to the center of the vortex at the south pole. The arrows on top of them show the corresponding heading direction of the microswimmer and their color refers to the starting angle (green: $\varphi_0=4.7122$, blue: $\varphi_0=4.7123$, red: $\varphi_0=4.7124$, yellow: $\varphi_0=4.7125$). In the initial part where they overlap, the path is highlighted in white. This shows a strong dependence on initial conditions for optimal trajectories passing close to the vortex at the south pole.
  \label{figiso}}
\vspace{-0.5truecm}
\end{figure*}

We can now directly compare the forward and backward paths in this setup, by showing how the area of the portion of sphere enclosed in the forward-backward loop varies with the amplitude of the external force, $A$. In Fig. \ref{fig1area}(b), we show the results obtained for one choice of initial and final points. 
The area enclosed in the loop grows as the amplitude of the force increases, which is expected since both paths deviate more from the Riemannian geodesic (the optimal path in the absence of external force). Interestingly, the enclosed area undergoes a jump when the vortex at the south pole is encircled, as beyond a certain threshold in the force amplitude the microswimmer can exploit the vortex to reach the goal more quickly and this causes an abrupt change in the shape of the optimal forward path. The scaling with $A$ is affected by this change, going from being linear (black dashed line) to sublinear.

\paragraph{Performance assessment.---}

We can now analyze the optimal paths obtained by following the Finsler geometry-based approach, which we call the \emph{Randers Policy} (RP), in comparison with a benchmark, which we refer to as the \emph{Direct Policy} (DP), in which the microswimmer always points in the direction of the target, regardless of the force field. 
To this end, we compute the time $T$ required to reach the target in units of the time $T_R$ it would take in the absence of any external force, as a function of the maximum force on the sphere. In Fig. \ref{figDP} we show the results obtained for two different choices of the initial and final points. 

In either case, for small values of the force, the two strategies do not show substantial differences in terms of performance. However, for the example shown on the left in Fig. \ref{figDP}, two particular situations can be observed. For larger values of the force (yellow and green regions) the RP (blue crosses) exploits the presence of the vortex at the south pole and at the same time the relative gain with respect to the DP (red circles) grows. In fact, following the former strategy makes it possible for the microswimmer to take up to $40\%$ less time to reach the target. Moreover, for sufficiently large values of the force intensity (green region), the DP also includes the vortex. This slightly helps the swimmer, although just for a small range of values (see the local minimum in the green region). In addition, the relative gain following the RP is substantial (up to about $20\%$ in terms of arrival time) even when this strategy does not imply the exploitation of any specific force field structures (see the plot on the right in Fig. \ref{figDP}).

\paragraph{Isochrone analysis.---}

To study more generally the behavior and the shape of the optimal trajectories coming from the RP, we analyze the so-called \emph{isochrones}, which are curves of equal travel time obtained by fixing the microswimmer initial position $(\theta_0,\phi_0,)$ and varying the starting angle $\varphi_0$ from 0 to $2\pi$. They can be seen as one-dimensional wavefronts of microswimmers that propagate onto the sphere following the Randers geodesics \eqref{eq:geodesic}. In the absence of a force field, the isochrones are concentric circles. 
In Fig. \ref{figiso}(a) we show some isochrones (solid lines) corresponding to the optimal paths starting from a point on the equator (green circle), in the presence of the force field with $A=\frac{1}{2}v_0$.

We observe that isochrones can feature self-intersections [see the example at $\tau=2.9\,v_0^{-1}$ on the right in Fig. \ref{figiso}(a)]. These are spots on the sphere for which there are multiple solutions to the problem of optimal navigation. 
Moreover, the isochrones can develop cusps, as highlighted in Fig. \ref{figiso}(a)  (see Supplemental Material \cite{SM}, Movie), which are points at which neighboring geodesics meet. These cusp are analogues of {\em conjugate points} in general relativity \cite{manor}, and related to the {\em caustics} in optics, as they represent domains on the isochrones with a higher density of geodesics \cite{arnold}.

The isochrones are considerably distorted after they encounter the vortex at the south pole [see the isochrone at $\tau=1.7\,v_0^{-1}$ in Fig. \ref{figiso}(a)]. The meaning of such deformation can be understood by looking at Fig. \ref{figiso}(b). Here are shown four optimal trajectories (dashed lines) starting from a point on the equator (green circle) and ending at time $\tau=2.2\,v_0^{-1}$ on the corresponding isochrone (blue solid line). Their initial angles $\varphi_0$ differ only by $\Delta\varphi_0=10^{-4}$. Such paths initially overlap (white dashed line) and separate only once they reach the south pole. Instead, all the optimal trajectories passing through the vortex at the north pole (which has a null intensity at its center) do not show this strong dependence on the initial conditions.


\paragraph{Concluding remarks.---}

We formulate and discuss a geometric description of the optimal navigation problem for microswimmers on curved manifolds. We show that this problem can be solved by finding the geodesics of a non-reversible Finsler metric of Randers type, providing a link between microswimmers physics and generalizations of general relativity.
Our proposed geometric approach provides tools for solving the optimal navigation problem in as yet unexplored, as well as more complex scenarios, such as paths for microswimmers escaping from harmful regions.

\begin{acknowledgments}
L.P. acknowledges A. Codutti for fruitful discussions. This work was supported by the Max Planck Society. 
\end{acknowledgments}

\bibliography{draft6}

\begin{thebibliography}{36}%
\makeatletter
\providecommand \@ifxundefined [1]{%
 \@ifx{#1\undefined}
}%
\providecommand \@ifnum [1]{%
 \ifnum #1\expandafter \@firstoftwo
 \else \expandafter \@secondoftwo
 \fi
}%
\providecommand \@ifx [1]{%
 \ifx #1\expandafter \@firstoftwo
 \else \expandafter \@secondoftwo
 \fi
}%
\providecommand \natexlab [1]{#1}%
\providecommand \enquote  [1]{``#1''}%
\providecommand \bibnamefont  [1]{#1}%
\providecommand \bibfnamefont [1]{#1}%
\providecommand \citenamefont [1]{#1}%
\providecommand \href@noop [0]{\@secondoftwo}%
\providecommand \href [0]{\begingroup \@sanitize@url \@href}%
\providecommand \@href[1]{\@@startlink{#1}\@@href}%
\providecommand \@@href[1]{\endgroup#1\@@endlink}%
\providecommand \@sanitize@url [0]{\catcode `\\12\catcode `\$12\catcode
  `\&12\catcode `\#12\catcode `\^12\catcode `\_12\catcode `\%12\relax}%
\providecommand \@@startlink[1]{}%
\providecommand \@@endlink[0]{}%
\providecommand \url  [0]{\begingroup\@sanitize@url \@url }%
\providecommand \@url [1]{\endgroup\@href {#1}{\urlprefix }}%
\providecommand \urlprefix  [0]{URL }%
\providecommand \Eprint [0]{\href }%
\providecommand \doibase [0]{http://dx.doi.org/}%
\providecommand \selectlanguage [0]{\@gobble}%
\providecommand \bibinfo  [0]{\@secondoftwo}%
\providecommand \bibfield  [0]{\@secondoftwo}%
\providecommand \translation [1]{[#1]}%
\providecommand \BibitemOpen [0]{}%
\providecommand \bibitemStop [0]{}%
\providecommand \bibitemNoStop [0]{.\EOS\space}%
\providecommand \EOS [0]{\spacefactor3000\relax}%
\providecommand \BibitemShut  [1]{\csname bibitem#1\endcsname}%
\let\auto@bib@innerbib\@empty
\bibitem [{\citenamefont {Berg}(2008)}]{berg}%
  \BibitemOpen
  \bibfield  {author} {\bibinfo {author} {\bibfnamefont {H.~C.}\ \bibnamefont
  {Berg}},\ }\href@noop {} {\emph {\bibinfo {title} {E. Coli in Motion}}}\
  (\bibinfo  {publisher} {Springer Science and Business Media},\ \bibinfo
  {year} {2008})\BibitemShut {NoStop}%
\bibitem [{\citenamefont {Bray}(2000)}]{Bray2000}%
  \BibitemOpen
  \bibfield  {author} {\bibinfo {author} {\bibfnamefont {D.}~\bibnamefont
  {Bray}},\ }\href {\doibase 10.4324/9780203833582} {\emph {\bibinfo {title}
  {Cell Movements}}}\ (\bibinfo  {publisher} {Garland Science},\ \bibinfo
  {year} {2000})\BibitemShut {NoStop}%
\bibitem [{\citenamefont {Bennett}\ and\ \citenamefont
  {Golestanian}(2015)}]{Bennett2015}%
  \BibitemOpen
  \bibfield  {author} {\bibinfo {author} {\bibfnamefont {R.~R.}\ \bibnamefont
  {Bennett}}\ and\ \bibinfo {author} {\bibfnamefont {R.}~\bibnamefont
  {Golestanian}},\ }\href {\doibase 10.1098/rsif.2014.1164} {\bibfield
  {journal} {\bibinfo  {journal} {Journal of The Royal Society Interface}\
  }\textbf {\bibinfo {volume} {12}},\ \bibinfo {pages} {20141164} (\bibinfo
  {year} {2015})}\BibitemShut {NoStop}%
\bibitem [{\citenamefont {Gompper}\ \emph {et~al.}(2020)\citenamefont
  {Gompper}, \citenamefont {Winkler}, \citenamefont {Speck}, \citenamefont
  {Solon}, \citenamefont {Nardini}, \citenamefont {Peruani}, \citenamefont
  {L{\"o}wen}, \citenamefont {Golestanian}, \citenamefont {Kaupp},
  \citenamefont {Alvarez}, \citenamefont {Ki{\o}rboe}, \citenamefont {Lauga},
  \citenamefont {Poon}, \citenamefont {DeSimone}, \citenamefont
  {Mui{\~{n}}os-Landin}, \citenamefont {Fischer}, \citenamefont {S{\"o}ker},
  \citenamefont {Cichos}, \citenamefont {Kapral}, \citenamefont {Gaspard},
  \citenamefont {Ripoll}, \citenamefont {Sagues}, \citenamefont
  {Doostmohammadi}, \citenamefont {Yeomans}, \citenamefont {Aranson},
  \citenamefont {Bechinger}, \citenamefont {Stark}, \citenamefont {Hemelrijk},
  \citenamefont {Nedelec}, \citenamefont {Sarkar}, \citenamefont {Aryaksama},
  \citenamefont {Lacroix}, \citenamefont {Duclos}, \citenamefont {Yashunsky},
  \citenamefont {Silberzan}, \citenamefont {Arroyo},\ and\ \citenamefont
  {Kale}}]{Gompper2020}%
  \BibitemOpen
  \bibfield  {author} {\bibinfo {author} {\bibfnamefont {G.}~\bibnamefont
  {Gompper}}, \bibinfo {author} {\bibfnamefont {R.~G.}\ \bibnamefont
  {Winkler}}, \bibinfo {author} {\bibfnamefont {T.}~\bibnamefont {Speck}},
  \bibinfo {author} {\bibfnamefont {A.}~\bibnamefont {Solon}}, \bibinfo
  {author} {\bibfnamefont {C.}~\bibnamefont {Nardini}}, \bibinfo {author}
  {\bibfnamefont {F.}~\bibnamefont {Peruani}}, \bibinfo {author} {\bibfnamefont
  {H.}~\bibnamefont {L{\"o}wen}}, \bibinfo {author} {\bibfnamefont
  {R.}~\bibnamefont {Golestanian}}, \bibinfo {author} {\bibfnamefont {U.~B.}\
  \bibnamefont {Kaupp}}, \bibinfo {author} {\bibfnamefont {L.}~\bibnamefont
  {Alvarez}}, \bibinfo {author} {\bibfnamefont {T.}~\bibnamefont {Ki{\o}rboe}},
  \bibinfo {author} {\bibfnamefont {E.}~\bibnamefont {Lauga}}, \bibinfo
  {author} {\bibfnamefont {W.~C.~K.}\ \bibnamefont {Poon}}, \bibinfo {author}
  {\bibfnamefont {A.}~\bibnamefont {DeSimone}}, \bibinfo {author}
  {\bibfnamefont {S.}~\bibnamefont {Mui{\~{n}}os-Landin}}, \bibinfo {author}
  {\bibfnamefont {A.}~\bibnamefont {Fischer}}, \bibinfo {author} {\bibfnamefont
  {N.~A.}\ \bibnamefont {S{\"o}ker}}, \bibinfo {author} {\bibfnamefont
  {F.}~\bibnamefont {Cichos}}, \bibinfo {author} {\bibfnamefont
  {R.}~\bibnamefont {Kapral}}, \bibinfo {author} {\bibfnamefont
  {P.}~\bibnamefont {Gaspard}}, \bibinfo {author} {\bibfnamefont
  {M.}~\bibnamefont {Ripoll}}, \bibinfo {author} {\bibfnamefont
  {F.}~\bibnamefont {Sagues}}, \bibinfo {author} {\bibfnamefont
  {A.}~\bibnamefont {Doostmohammadi}}, \bibinfo {author} {\bibfnamefont
  {J.~M.}\ \bibnamefont {Yeomans}}, \bibinfo {author} {\bibfnamefont {I.~S.}\
  \bibnamefont {Aranson}}, \bibinfo {author} {\bibfnamefont {C.}~\bibnamefont
  {Bechinger}}, \bibinfo {author} {\bibfnamefont {H.}~\bibnamefont {Stark}},
  \bibinfo {author} {\bibfnamefont {C.~K.}\ \bibnamefont {Hemelrijk}}, \bibinfo
  {author} {\bibfnamefont {F.~J.}\ \bibnamefont {Nedelec}}, \bibinfo {author}
  {\bibfnamefont {T.}~\bibnamefont {Sarkar}}, \bibinfo {author} {\bibfnamefont
  {T.}~\bibnamefont {Aryaksama}}, \bibinfo {author} {\bibfnamefont
  {M.}~\bibnamefont {Lacroix}}, \bibinfo {author} {\bibfnamefont
  {G.}~\bibnamefont {Duclos}}, \bibinfo {author} {\bibfnamefont
  {V.}~\bibnamefont {Yashunsky}}, \bibinfo {author} {\bibfnamefont
  {P.}~\bibnamefont {Silberzan}}, \bibinfo {author} {\bibfnamefont
  {M.}~\bibnamefont {Arroyo}}, \ and\ \bibinfo {author} {\bibfnamefont
  {S.}~\bibnamefont {Kale}},\ }\href {\doibase 10.1088/1361-648x/ab6348}
  {\bibfield  {journal} {\bibinfo  {journal} {Journal of Physics: Condensed
  Matter}\ }\textbf {\bibinfo {volume} {32}},\ \bibinfo {pages} {193001}
  (\bibinfo {year} {2020})}\BibitemShut {NoStop}%
\bibitem [{\citenamefont {Golestanian}\ \emph {et~al.}(2007)\citenamefont
  {Golestanian}, \citenamefont {Liverpool},\ and\ \citenamefont
  {Ajdari}}]{Golestanian2007}%
  \BibitemOpen
  \bibfield  {author} {\bibinfo {author} {\bibfnamefont {R.}~\bibnamefont
  {Golestanian}}, \bibinfo {author} {\bibfnamefont {T.~B.}\ \bibnamefont
  {Liverpool}}, \ and\ \bibinfo {author} {\bibfnamefont {A.}~\bibnamefont
  {Ajdari}},\ }\href {\doibase 10.1088/1367-2630/9/5/126} {\bibfield  {journal}
  {\bibinfo  {journal} {New Journal of Physics}\ }\textbf {\bibinfo {volume}
  {9}},\ \bibinfo {pages} {126} (\bibinfo {year} {2007})}\BibitemShut {NoStop}%
\bibitem [{\citenamefont {Golestanian}(2019)}]{RG-phoretic}%
  \BibitemOpen
  \bibfield  {author} {\bibinfo {author} {\bibfnamefont {R.}~\bibnamefont
  {Golestanian}},\ }\href@noop {} {\enquote {\bibinfo {title} {Phoretic active
  matter arxiv:1909.03747},}\ } (\bibinfo {year} {2019}),\ \Eprint
  {http://arxiv.org/abs/arXiv:1909.03747} {arXiv:1909.03747} \BibitemShut
  {NoStop}%
\bibitem [{\citenamefont {Stark}(2018)}]{stark}%
  \BibitemOpen
  \bibfield  {author} {\bibinfo {author} {\bibfnamefont {H.}~\bibnamefont
  {Stark}},\ }\href@noop {} {\bibfield  {journal} {\bibinfo  {journal} {Acc.
  Chem. Res.}\ }\textbf {\bibinfo {volume} {51}},\ \bibinfo {pages} {2355}
  (\bibinfo {year} {2018})}\BibitemShut {NoStop}%
\bibitem [{\citenamefont {Mano}\ \emph {et~al.}(2017)\citenamefont {Mano},
  \citenamefont {Delfau}, \citenamefont {Iwasawa},\ and\ \citenamefont
  {Sano}}]{mano}%
  \BibitemOpen
  \bibfield  {author} {\bibinfo {author} {\bibfnamefont {T.}~\bibnamefont
  {Mano}}, \bibinfo {author} {\bibfnamefont {J.-B.}\ \bibnamefont {Delfau}},
  \bibinfo {author} {\bibfnamefont {J.}~\bibnamefont {Iwasawa}}, \ and\
  \bibinfo {author} {\bibfnamefont {M.}~\bibnamefont {Sano}},\ }\href@noop {}
  {\bibfield  {journal} {\bibinfo  {journal} {Proc. Natl. Acad. Sci.}\ }\textbf
  {\bibinfo {volume} {114}},\ \bibinfo {pages} {2580} (\bibinfo {year}
  {2017})}\BibitemShut {NoStop}%
\bibitem [{\citenamefont {Tierno}\ \emph {et~al.}(2008)\citenamefont {Tierno},
  \citenamefont {Golestanian}, \citenamefont {Pagonabarraga},\ and\
  \citenamefont {Sagues}}]{Tierno2008b}%
  \BibitemOpen
  \bibfield  {author} {\bibinfo {author} {\bibfnamefont {P.}~\bibnamefont
  {Tierno}}, \bibinfo {author} {\bibfnamefont {R.}~\bibnamefont {Golestanian}},
  \bibinfo {author} {\bibfnamefont {I.}~\bibnamefont {Pagonabarraga}}, \ and\
  \bibinfo {author} {\bibfnamefont {F.}~\bibnamefont {Sagues}},\ }\href
  {\doibase 10.1021/jp808354n} {\bibfield  {journal} {\bibinfo  {journal} {The
  Journal of Physical Chemistry B}\ }\textbf {\bibinfo {volume} {112}},\
  \bibinfo {pages} {16525} (\bibinfo {year} {2008})}\BibitemShut {NoStop}%
\bibitem [{\citenamefont {Bauerle}\ \emph {et~al.}(2018)\citenamefont
  {Bauerle}, \citenamefont {Fischer}, \citenamefont {Speck},\ and\
  \citenamefont {Bechinger}}]{bauerle}%
  \BibitemOpen
  \bibfield  {author} {\bibinfo {author} {\bibfnamefont {T.}~\bibnamefont
  {Bauerle}}, \bibinfo {author} {\bibfnamefont {A.}~\bibnamefont {Fischer}},
  \bibinfo {author} {\bibfnamefont {T.}~\bibnamefont {Speck}}, \ and\ \bibinfo
  {author} {\bibfnamefont {C.}~\bibnamefont {Bechinger}},\ }\href@noop {}
  {\bibfield  {journal} {\bibinfo  {journal} {Nature Comm.}\ }\textbf {\bibinfo
  {volume} {9}},\ \bibinfo {pages} {3232} (\bibinfo {year} {2018})}\BibitemShut
  {NoStop}%
\bibitem [{\citenamefont {Khadka}\ \emph {et~al.}(2018)\citenamefont {Khadka},
  \citenamefont {Holubec}, \citenamefont {Yang},\ and\ \citenamefont
  {Cichos}}]{khadka}%
  \BibitemOpen
  \bibfield  {author} {\bibinfo {author} {\bibfnamefont {U.}~\bibnamefont
  {Khadka}}, \bibinfo {author} {\bibfnamefont {V.}~\bibnamefont {Holubec}},
  \bibinfo {author} {\bibfnamefont {H.}~\bibnamefont {Yang}}, \ and\ \bibinfo
  {author} {\bibfnamefont {F.}~\bibnamefont {Cichos}},\ }\href@noop {}
  {\bibfield  {journal} {\bibinfo  {journal} {Nature Comm.}\ }\textbf {\bibinfo
  {volume} {9}},\ \bibinfo {pages} {3864} (\bibinfo {year} {2018})}\BibitemShut
  {NoStop}%
\bibitem [{\citenamefont {Das}\ \emph {et~al.}(2015)\citenamefont {Das},
  \citenamefont {Garg}, \citenamefont {Campbell}, \citenamefont {Howse},
  \citenamefont {Sen}, \citenamefont {Velegol}, \citenamefont {Golestanian},\
  and\ \citenamefont {Ebbens}}]{Das2015}%
  \BibitemOpen
  \bibfield  {author} {\bibinfo {author} {\bibfnamefont {S.}~\bibnamefont
  {Das}}, \bibinfo {author} {\bibfnamefont {A.}~\bibnamefont {Garg}}, \bibinfo
  {author} {\bibfnamefont {A.~I.}\ \bibnamefont {Campbell}}, \bibinfo {author}
  {\bibfnamefont {J.}~\bibnamefont {Howse}}, \bibinfo {author} {\bibfnamefont
  {A.}~\bibnamefont {Sen}}, \bibinfo {author} {\bibfnamefont {D.}~\bibnamefont
  {Velegol}}, \bibinfo {author} {\bibfnamefont {R.}~\bibnamefont
  {Golestanian}}, \ and\ \bibinfo {author} {\bibfnamefont {S.~J.}\ \bibnamefont
  {Ebbens}},\ }\href {\doibase 10.1038/ncomms9999} {\bibfield  {journal}
  {\bibinfo  {journal} {Nature Communications}\ }\textbf {\bibinfo {volume}
  {6}} (\bibinfo {year} {2015}),\ 10.1038/ncomms9999}\BibitemShut {NoStop}%
\bibitem [{\citenamefont {Park}\ \emph {et~al.}(2017)\citenamefont {Park},
  \citenamefont {Zhuang}, \citenamefont {Yasa},\ and\ \citenamefont
  {Sitti}}]{park}%
  \BibitemOpen
  \bibfield  {author} {\bibinfo {author} {\bibfnamefont {B.-W.}\ \bibnamefont
  {Park}}, \bibinfo {author} {\bibfnamefont {J.}~\bibnamefont {Zhuang}},
  \bibinfo {author} {\bibfnamefont {O.}~\bibnamefont {Yasa}}, \ and\ \bibinfo
  {author} {\bibfnamefont {M.}~\bibnamefont {Sitti}},\ }\href@noop {}
  {\bibfield  {journal} {\bibinfo  {journal} {ACS Nano}\ }\textbf {\bibinfo
  {volume} {11}},\ \bibinfo {pages} {8910} (\bibinfo {year}
  {2017})}\BibitemShut {NoStop}%
\bibitem [{\citenamefont {Qiu}\ \emph {et~al.}(2015)\citenamefont {Qiu},
  \citenamefont {Fujita}, \citenamefont {Mhanna}, \citenamefont {Zhang},
  \citenamefont {Simona},\ and\ \citenamefont {Nelson}}]{qiu}%
  \BibitemOpen
  \bibfield  {author} {\bibinfo {author} {\bibfnamefont {F.}~\bibnamefont
  {Qiu}}, \bibinfo {author} {\bibfnamefont {S.}~\bibnamefont {Fujita}},
  \bibinfo {author} {\bibfnamefont {R.}~\bibnamefont {Mhanna}}, \bibinfo
  {author} {\bibfnamefont {L.}~\bibnamefont {Zhang}}, \bibinfo {author}
  {\bibfnamefont {B.~R.}\ \bibnamefont {Simona}}, \ and\ \bibinfo {author}
  {\bibfnamefont {B.~J.}\ \bibnamefont {Nelson}},\ }\href@noop {} {\bibfield
  {journal} {\bibinfo  {journal} {Adv. Func. Mater.}\ }\textbf {\bibinfo
  {volume} {25}},\ \bibinfo {pages} {1666} (\bibinfo {year}
  {2015})}\BibitemShut {NoStop}%
\bibitem [{\citenamefont {Demirörs}\ \emph {et~al.}(2018)\citenamefont
  {Demirörs}, \citenamefont {Akan}, \citenamefont {Poloni},\ and\
  \citenamefont {Studart}}]{demirors}%
  \BibitemOpen
  \bibfield  {author} {\bibinfo {author} {\bibfnamefont {A.~F.}\ \bibnamefont
  {Demirörs}}, \bibinfo {author} {\bibfnamefont {M.~T.}\ \bibnamefont {Akan}},
  \bibinfo {author} {\bibfnamefont {E.}~\bibnamefont {Poloni}}, \ and\ \bibinfo
  {author} {\bibfnamefont {A.~R.}\ \bibnamefont {Studart}},\ }\href {\doibase
  10.1039/C8SM00513C} {\bibfield  {journal} {\bibinfo  {journal} {Soft Matter}\
  }\textbf {\bibinfo {volume} {14}},\ \bibinfo {pages} {4741} (\bibinfo {year}
  {2018})}\BibitemShut {NoStop}%
\bibitem [{\citenamefont {Villa}\ \emph {et~al.}(2020)\citenamefont {Villa},
  \citenamefont {Viktorova}, \citenamefont {Plutnar}, \citenamefont {Ruml},
  \citenamefont {Hoang},\ and\ \citenamefont {Pumera}}]{Villa2020}%
  \BibitemOpen
  \bibfield  {author} {\bibinfo {author} {\bibfnamefont {K.}~\bibnamefont
  {Villa}}, \bibinfo {author} {\bibfnamefont {J.}~\bibnamefont {Viktorova}},
  \bibinfo {author} {\bibfnamefont {J.}~\bibnamefont {Plutnar}}, \bibinfo
  {author} {\bibfnamefont {T.}~\bibnamefont {Ruml}}, \bibinfo {author}
  {\bibfnamefont {L.}~\bibnamefont {Hoang}}, \ and\ \bibinfo {author}
  {\bibfnamefont {M.}~\bibnamefont {Pumera}},\ }\href {\doibase
  10.1016/j.xcrp.2020.100181} {\bibfield  {journal} {\bibinfo  {journal} {Cell
  Reports Physical Science}\ }\textbf {\bibinfo {volume} {1}},\ \bibinfo
  {pages} {100181} (\bibinfo {year} {2020})}\BibitemShut {NoStop}%
\bibitem [{\citenamefont {Zermelo}(1931)}]{zermelo}%
  \BibitemOpen
  \bibfield  {author} {\bibinfo {author} {\bibfnamefont {E.}~\bibnamefont
  {Zermelo}},\ }\href@noop {} {\bibfield  {journal} {\bibinfo  {journal} {Math.
  Phys.}\ }\textbf {\bibinfo {volume} {11}},\ \bibinfo {pages} {114} (\bibinfo
  {year} {1931})}\BibitemShut {NoStop}%
\bibitem [{\citenamefont {Liebchen}\ and\ \citenamefont
  {Lowen}(2019)}]{liebchen}%
  \BibitemOpen
  \bibfield  {author} {\bibinfo {author} {\bibfnamefont {B.}~\bibnamefont
  {Liebchen}}\ and\ \bibinfo {author} {\bibfnamefont {H.}~\bibnamefont
  {Lowen}},\ }\href@noop {} {\bibfield  {journal} {\bibinfo  {journal} {EPL}\
  }\textbf {\bibinfo {volume} {127}},\ \bibinfo {pages} {3} (\bibinfo {year}
  {2019})}\BibitemShut {NoStop}%
\bibitem [{\citenamefont {Colabrese}\ \emph {et~al.}(2017)\citenamefont
  {Colabrese}, \citenamefont {Gustavsson}, \citenamefont {Celani},\ and\
  \citenamefont {Biferale}}]{colabrese}%
  \BibitemOpen
  \bibfield  {author} {\bibinfo {author} {\bibfnamefont {S.}~\bibnamefont
  {Colabrese}}, \bibinfo {author} {\bibfnamefont {K.}~\bibnamefont
  {Gustavsson}}, \bibinfo {author} {\bibfnamefont {A.}~\bibnamefont {Celani}},
  \ and\ \bibinfo {author} {\bibfnamefont {L.}~\bibnamefont {Biferale}},\
  }\href {\doibase 10.1103/PhysRevLett.118.158004} {\bibfield  {journal}
  {\bibinfo  {journal} {Phys. Rev. Lett.}\ }\textbf {\bibinfo {volume} {118}},\
  \bibinfo {pages} {158004} (\bibinfo {year} {2017})}\BibitemShut {NoStop}%
\bibitem [{\citenamefont {Biferale}\ \emph {et~al.}(2019)\citenamefont
  {Biferale}, \citenamefont {Bonaccorso}, \citenamefont {Buzzicotti},
  \citenamefont {Leoni},\ and\ \citenamefont {Gustavsson}}]{biferale}%
  \BibitemOpen
  \bibfield  {author} {\bibinfo {author} {\bibfnamefont {L.}~\bibnamefont
  {Biferale}}, \bibinfo {author} {\bibfnamefont {F.}~\bibnamefont
  {Bonaccorso}}, \bibinfo {author} {\bibfnamefont {M.}~\bibnamefont
  {Buzzicotti}}, \bibinfo {author} {\bibfnamefont {P.~C.~D.}\ \bibnamefont
  {Leoni}}, \ and\ \bibinfo {author} {\bibfnamefont {K.}~\bibnamefont
  {Gustavsson}},\ }\href@noop {} {\bibfield  {journal} {\bibinfo  {journal}
  {Chaos}\ }\textbf {\bibinfo {volume} {29}},\ \bibinfo {pages} {103138}
  (\bibinfo {year} {2019})}\BibitemShut {NoStop}%
\bibitem [{\citenamefont {Schneider}\ and\ \citenamefont
  {Stark}(2019)}]{schneider}%
  \BibitemOpen
  \bibfield  {author} {\bibinfo {author} {\bibfnamefont {E.}~\bibnamefont
  {Schneider}}\ and\ \bibinfo {author} {\bibfnamefont {H.}~\bibnamefont
  {Stark}},\ }\href@noop {} {\bibfield  {journal} {\bibinfo  {journal} {EPL}\
  }\textbf {\bibinfo {volume} {127}},\ \bibinfo {pages} {6} (\bibinfo {year}
  {2019})}\BibitemShut {NoStop}%
\bibitem [{\citenamefont {Shen}(2003)}]{shen}%
  \BibitemOpen
  \bibfield  {author} {\bibinfo {author} {\bibfnamefont {Z.}~\bibnamefont
  {Shen}},\ }\href@noop {} {\bibfield  {journal} {\bibinfo  {journal} {Canadian
  J. Math.}\ }\textbf {\bibinfo {volume} {55}},\ \bibinfo {pages} {112}
  (\bibinfo {year} {2003})}\BibitemShut {NoStop}%
\bibitem [{\citenamefont {Bao}\ \emph {et~al.}(2004)\citenamefont {Bao},
  \citenamefont {Robles},\ and\ \citenamefont {Shen}}]{bao}%
  \BibitemOpen
  \bibfield  {author} {\bibinfo {author} {\bibfnamefont {D.}~\bibnamefont
  {Bao}}, \bibinfo {author} {\bibfnamefont {C.}~\bibnamefont {Robles}}, \ and\
  \bibinfo {author} {\bibfnamefont {Z.}~\bibnamefont {Shen}},\ }\href@noop {}
  {\bibfield  {journal} {\bibinfo  {journal} {J. Diff. Geom.}\ }\textbf
  {\bibinfo {volume} {66}},\ \bibinfo {pages} {377} (\bibinfo {year}
  {2004})}\BibitemShut {NoStop}%
\bibitem [{Fin()}]{Finsler}%
  \BibitemOpen
  \href@noop {} {}\bibinfo {note} {P. {Finsler}, \"Uber Kurven und Fl\"achen in
  allgemeinen R\"aumen, Dissertation, University of G\"ottingen
  (1918).}\BibitemShut {Stop}%
\bibitem [{\citenamefont {Randers}(1941)}]{randers}%
  \BibitemOpen
  \bibfield  {author} {\bibinfo {author} {\bibfnamefont {G.}~\bibnamefont
  {Randers}},\ }\href@noop {} {\bibfield  {journal} {\bibinfo  {journal} {Phys.
  Rev.}\ }\textbf {\bibinfo {volume} {59}},\ \bibinfo {pages} {195} (\bibinfo
  {year} {1941})}\BibitemShut {NoStop}%
\bibitem [{\citenamefont {Gibbons}\ \emph {et~al.}(2009)\citenamefont
  {Gibbons}, \citenamefont {Herdeiro}, \citenamefont {Warnick},\ and\
  \citenamefont {Werner}}]{gibbons}%
  \BibitemOpen
  \bibfield  {author} {\bibinfo {author} {\bibfnamefont {G.~W.}\ \bibnamefont
  {Gibbons}}, \bibinfo {author} {\bibfnamefont {C.~A.~R.}\ \bibnamefont
  {Herdeiro}}, \bibinfo {author} {\bibfnamefont {C.~M.}\ \bibnamefont
  {Warnick}}, \ and\ \bibinfo {author} {\bibfnamefont {M.~C.}\ \bibnamefont
  {Werner}},\ }\href {\doibase 10.1103/PhysRevD.79.044022} {\bibfield
  {journal} {\bibinfo  {journal} {Phys. Rev. D}\ }\textbf {\bibinfo {volume}
  {79}},\ \bibinfo {pages} {044022} (\bibinfo {year} {2009})}\BibitemShut
  {NoStop}%
\bibitem [{\citenamefont {Brody}\ and\ \citenamefont {Meier}(2015)}]{brody}%
  \BibitemOpen
  \bibfield  {author} {\bibinfo {author} {\bibfnamefont {D.~C.}\ \bibnamefont
  {Brody}}\ and\ \bibinfo {author} {\bibfnamefont {D.~M.}\ \bibnamefont
  {Meier}},\ }\href {\doibase 10.1103/PhysRevLett.114.100502} {\bibfield
  {journal} {\bibinfo  {journal} {Phys. Rev. Lett.}\ }\textbf {\bibinfo
  {volume} {114}},\ \bibinfo {pages} {100502} (\bibinfo {year}
  {2015})}\BibitemShut {NoStop}%
\bibitem [{\citenamefont {Golestanian}\ \emph {et~al.}(1995)\citenamefont
  {Golestanian}, \citenamefont {Khajehpour},\ and\ \citenamefont
  {Mansouri}}]{Golestanian1995}%
  \BibitemOpen
  \bibfield  {author} {\bibinfo {author} {\bibfnamefont {R.}~\bibnamefont
  {Golestanian}}, \bibinfo {author} {\bibfnamefont {M.~R.~H.}\ \bibnamefont
  {Khajehpour}}, \ and\ \bibinfo {author} {\bibfnamefont {R.}~\bibnamefont
  {Mansouri}},\ }\href {\doibase 10.1088/0264-9381/12/1/021} {\bibfield
  {journal} {\bibinfo  {journal} {Classical and Quantum Gravity}\ }\textbf
  {\bibinfo {volume} {12}},\ \bibinfo {pages} {273} (\bibinfo {year}
  {1995})}\BibitemShut {NoStop}%
\bibitem [{\citenamefont {Schutz}(1980)}]{schutz}%
  \BibitemOpen
  \bibfield  {author} {\bibinfo {author} {\bibfnamefont {B.}~\bibnamefont
  {Schutz}},\ }\href@noop {} {\emph {\bibinfo {title} {Geometrical Methods of
  Mathematical Physics}}}\ (\bibinfo  {publisher} {Cambridge University
  Press},\ \bibinfo {year} {1980})\BibitemShut {NoStop}%
\bibitem [{\citenamefont {Bao}\ \emph {et~al.}(2000)\citenamefont {Bao},
  \citenamefont {Chern},\ and\ \citenamefont {Shen}}]{bao2}%
  \BibitemOpen
  \bibfield  {author} {\bibinfo {author} {\bibfnamefont {D.}~\bibnamefont
  {Bao}}, \bibinfo {author} {\bibfnamefont {S.~S.}\ \bibnamefont {Chern}}, \
  and\ \bibinfo {author} {\bibfnamefont {Z.}~\bibnamefont {Shen}},\ }\href@noop
  {} {\emph {\bibinfo {title} {An Introduction to Riemann-Finsler Geometry}}}\
  (\bibinfo  {publisher} {Springer, New York, NY},\ \bibinfo {year}
  {2000})\BibitemShut {NoStop}%
\bibitem [{\citenamefont {Chern}\ and\ \citenamefont {Shen}(2005)}]{chern}%
  \BibitemOpen
  \bibfield  {author} {\bibinfo {author} {\bibfnamefont {S.~S.}\ \bibnamefont
  {Chern}}\ and\ \bibinfo {author} {\bibfnamefont {Z.}~\bibnamefont {Shen}},\
  }\href@noop {} {\emph {\bibinfo {title} {Riemann-Finsler Geometry}}}\
  (\bibinfo  {publisher} {World Scientific},\ \bibinfo {year}
  {2005})\BibitemShut {NoStop}%
\bibitem [{\citenamefont {Cheng}\ and\ \citenamefont {Shen}(2012)}]{cheng}%
  \BibitemOpen
  \bibfield  {author} {\bibinfo {author} {\bibfnamefont {X.}~\bibnamefont
  {Cheng}}\ and\ \bibinfo {author} {\bibfnamefont {Z.}~\bibnamefont {Shen}},\
  }\href@noop {} {\emph {\bibinfo {title} {Finsler Geometry: An Approach Via
  Randers Spaces}}}\ (\bibinfo  {publisher} {Springer},\ \bibinfo {year}
  {2012})\BibitemShut {NoStop}%
\bibitem [{\citenamefont {Lee}(1997)}]{lee}%
  \BibitemOpen
  \bibfield  {author} {\bibinfo {author} {\bibfnamefont {J.~M.}\ \bibnamefont
  {Lee}},\ }\enquote {\bibinfo {title} {Riemannian geodesics},}\ in\ \href
  {\doibase 10.1007/0-387-22726-1_5} {\emph {\bibinfo {booktitle} {Riemannian
  Manifolds: An Introduction to Curvature}}}\ (\bibinfo  {publisher} {Springer
  New York},\ \bibinfo {address} {New York, NY},\ \bibinfo {year} {1997})\ pp.\
  \bibinfo {pages} {65--89}\BibitemShut {NoStop}%
\bibitem [{SM()}]{SM}%
  \BibitemOpen
  \href@noop {} {}\bibinfo {note} {See Supplemental Material at [...] for the
  Movie that shows the degeneracy of the optimization problem at the
  cusps.}\BibitemShut {Stop}%
\bibitem [{\citenamefont {Manor}(1977)}]{manor}%
  \BibitemOpen
  \bibfield  {author} {\bibinfo {author} {\bibfnamefont {Y.}~\bibnamefont
  {Manor}},\ }\href@noop {} {\bibfield  {journal} {\bibinfo  {journal} {Ann.
  Phys. (N.Y.)}\ }\textbf {\bibinfo {volume} {106}},\ \bibinfo {pages} {407}
  (\bibinfo {year} {1977})}\BibitemShut {NoStop}%
\bibitem [{\citenamefont {Arnold}(1990)}]{arnold}%
  \BibitemOpen
  \bibfield  {author} {\bibinfo {author} {\bibfnamefont {V.}~\bibnamefont
  {Arnold}},\ }\href@noop {} {\emph {\bibinfo {title} {Singularities of
  Caustics and Wave Fronts}}}\ (\bibinfo  {publisher} {Springer Netherlands},\
  \bibinfo {year} {1990})\BibitemShut {NoStop}%
\end{thebibliography}%

\end{document}